\documentclass{blois}

\def\be{\begin{equation}}
\def\ee{\end{equation}}
\def\bea{\begin{eqnarray}}
\def\eea{\end{eqnarray}}

\newcommand{\Photo}{}

\begin{document}
\vspace*{4cm}
\title{Interplay between Parton Distribution Functions and New Physics signals}

\author{ Elie Hammou }

\address{DAMTP, University of Cambridge, Wilberforce Road, Cambridge, CB3 0WA, United Kingdom}

\maketitle
\abstracts{
The analysis and the interpretation of the LHC data require a precise determination of Parton Distribution Functions (PDFs) in order to detect reliably potential signs of new physics. I present a systematic study designed to assess the risk of absorbing, and thus missing, signals from heavy new physics in the PDFs parameterisation during the High-Luminosity LHC run. I discuss the consequences of such a PDF "contamination" and consider possible solutions to it, for example the inclusion of other experimental data probing large-x regions at low-energy.}

\section{Introduction}

PDFs are used to compute all theoretical predictions at hadron colliders. They describe the repartition on the proton momentum among its constitutive partons (quarks and gluons). Their dependence on the energy scale $Q$ can be theoretically described with the DGLAP evolution equation, However, their dependence on the Bjorken $x$ cannot be predicted perturbatively, it has to be fitted from data \cite{SA} \cite{JG} \cite{KK}.

This study is assessing whether these fits could potentially absorb within the PDFs higher energies signals associated to new physics. The considered scenario is one where the SM Lagrangian is extended by additional terms corresponding to a BSM heavy field. Such contributions would have an impact on the high energy tails of some observables. If this data is used to perform a PDF fit assuming the SM in the theory predictions, the new physics could be "fitted away" provided the PDFs had sufficient degrees of freedom to adapt to the BSM induced shift without worsening the data-theory agreement of the other dataset included in the fit. The complete study can be found in a recent publication ~\cite{PBSP}.

\section{Strategy for assessing the risk of PDF contamination}

\subsection{General methodology}

In practice, this study relies on the use the "closure test" method developed by the NNPDF collaboration \cite{NNPDF}. In a nutshell, this is a three-step procedure. First, one chooses a PDF set that is assumed to be the "true" description of the proton structure. I will refer to it as the "initial PDFs". Second, one generates pseudodata with Monte Carlo methods by convoluting the initial PDFs with partonic cross sections computed perturbatively from a chosen Lagrangian. Third and last step, one performs a PDF fit on the obtained pseudodata. The output PDFs are then compared to the initial ones. If they are compatible with each other, one can assume that the fitting method is sound.

In this study, the method I just presented is modified slightly. Two types of closure tests are done. One in which the Lagrangian used to produce the pseudodata in the second step is the SM one, I will refer to this procedure as "baseline fit" and to its result as a "baseline PDFs". The other is performed with a Lagrangian containing additional new heavy fields, detailed below, in the second step. The fundamental point, is that in the third step, the PDFs are fitted only assuming the SM, as we do in real life from experimental data. This type of fit will be referred to as "contaminated fit" outputting "contaminated PDFs". Furthermore, the comparison here is between the baseline and contaminated PDFs. In the case that they are not compatible with each other, it would mean that the new physics has been fitted away in the contaminated PDFs.

\subsection{New heavy physics used for pseudodata generation}

To create the pseudodata, two new physics scenarios have been used, each involving a new heavy boson. One involving a heavy $Z'$ charged under the gauge group $U(1)_Y$ and the other a heavy $W'$ charged under $SU(2)_L$. Both these UV-complete models are matched to SM Effective Field Theory (SMEFT) Lagrangians with terms up to dimension six. This allows more flexibility in the choice of the parameters values.

\begin{figure}[h!]
    \centering
    \includegraphics[width=0.49\linewidth]{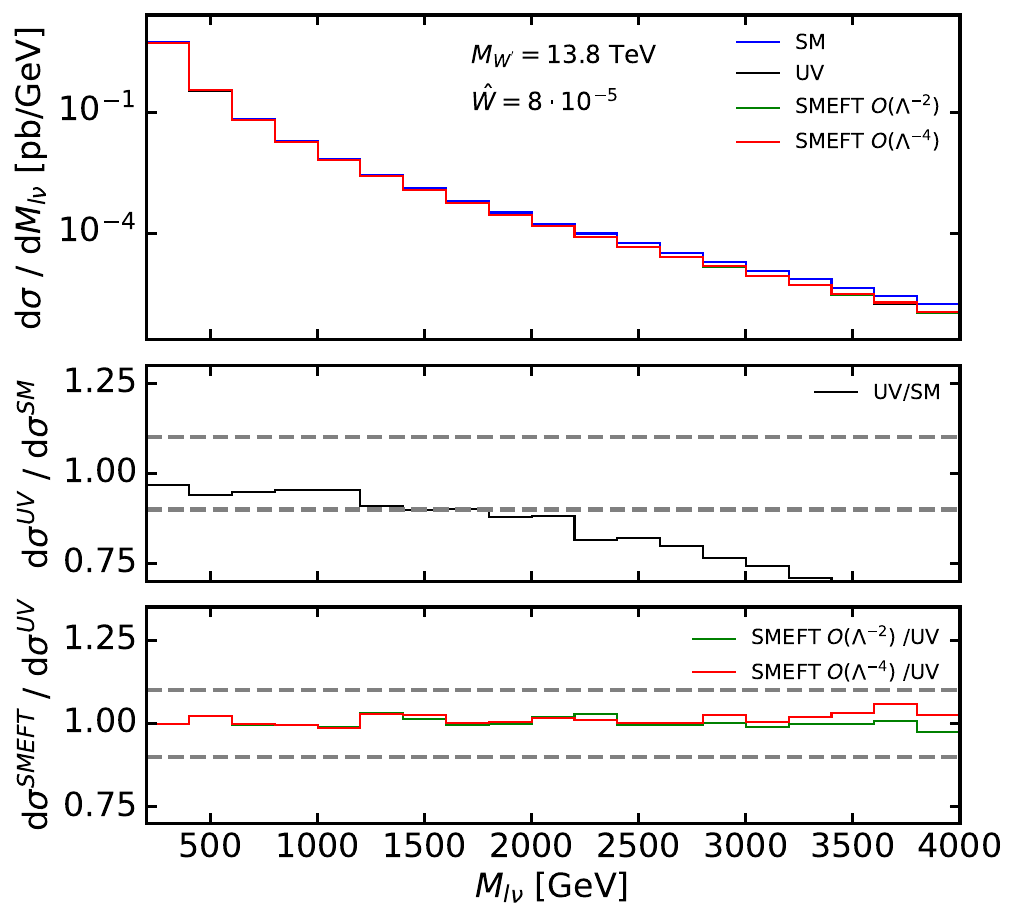}
    \includegraphics[width=0.49\linewidth]{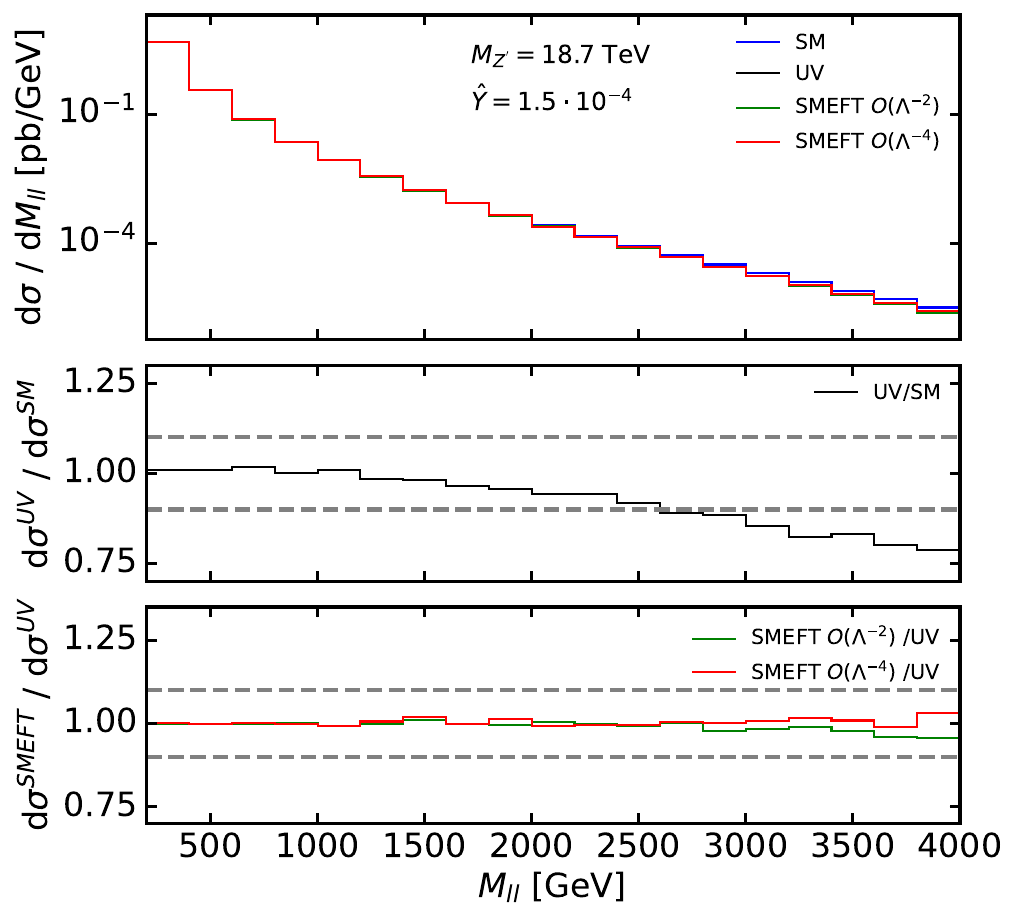}
    \caption{Predictions for Drell-Yan differential cross sections in dilepton invariant mass: $p \bar{p} \rightarrow l^- \bar{\nu}$ with a $W'$ on the left and $p \bar{p} \rightarrow l^- l^+$ with a $Z'$ on the right.}
    \label{fig:UV_EFT_comp}
\end{figure}

Then, the models are used to generate Drell-Yan (DY) pseudodata. The comparison of the predictions of the SM, the UV BSM, the SMEFT with only linear corrections and the SMEFT with both linear and quadratics can be seen in Fig. \ref{fig:UV_EFT_comp}.

As one can see, the middle panels show that in both cases the new physics introduced has a non trivial effect on the high energy tail of the distributions. The bottom panels show that in both cases the SMEFT models including only the linear corrections describe faithfully the UV physics up to 4 TeV. Those are the models that are used to generate the MC pseudodata which are added to the PDFs fits.

\section{Contamination from Drell-Yan large invariant masses distributions}

\subsection{Effect of new heavy bosons in PDF fits}

In the $Z'$ scenario, addition of the contaminated DY pseudodata worsened the quality of the fit. The PDFs were sufficiently constrained by the rest of the datasets to be unable to adapt to the BSM shift, resulting in a poor data-theory agreement for the DY processes. Consequently, the contaminated pseudodata was dropped from the fit and the baseline and contaminated PDFs were compatible with one another. No actual contamination has occurred.

However, in the $W'$ scenario the addition of the contaminated pseudodata has not lowered the quality of the fit, allowing it to be properly included in it. The baseline and contaminated PDFs were not compatible with each other suggesting that the impact of the $W'$ had been absorbed by the contaminated PDFs. This is due to the lack of constraints on the large-x antiquark distributions from the rest of the data included in the analysis. As a result, we would risk to miss the new physics while analysing this data.

\subsection{Impact of contamination on other observables}

\begin{figure}[h!]
    \centering
    \includegraphics[width=0.47\linewidth]{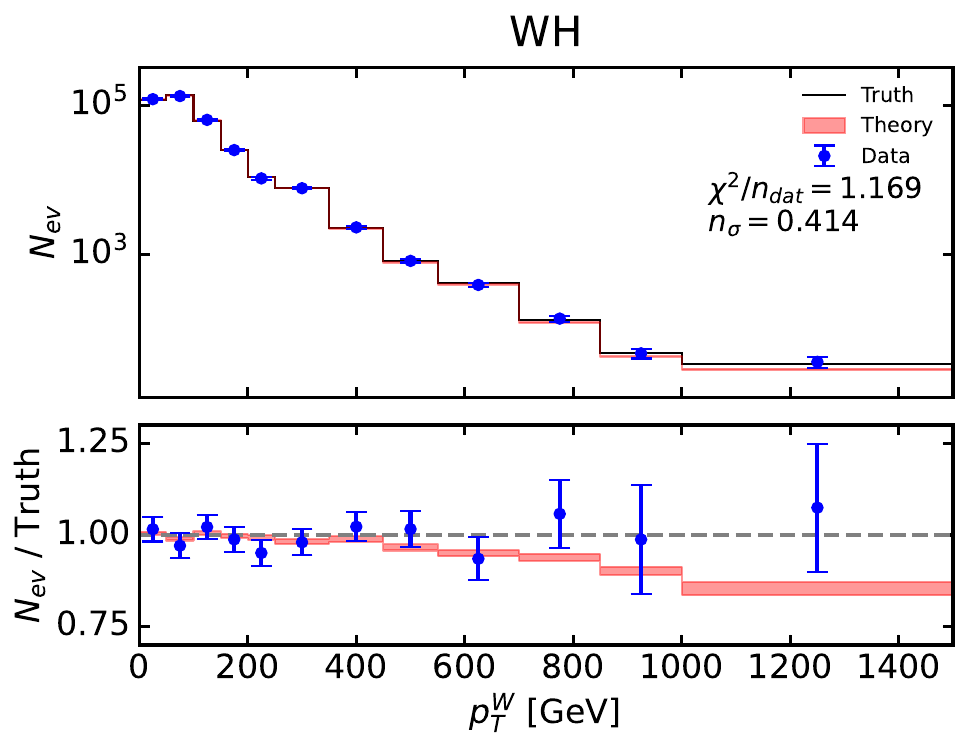}
    \includegraphics[width=0.49\linewidth]{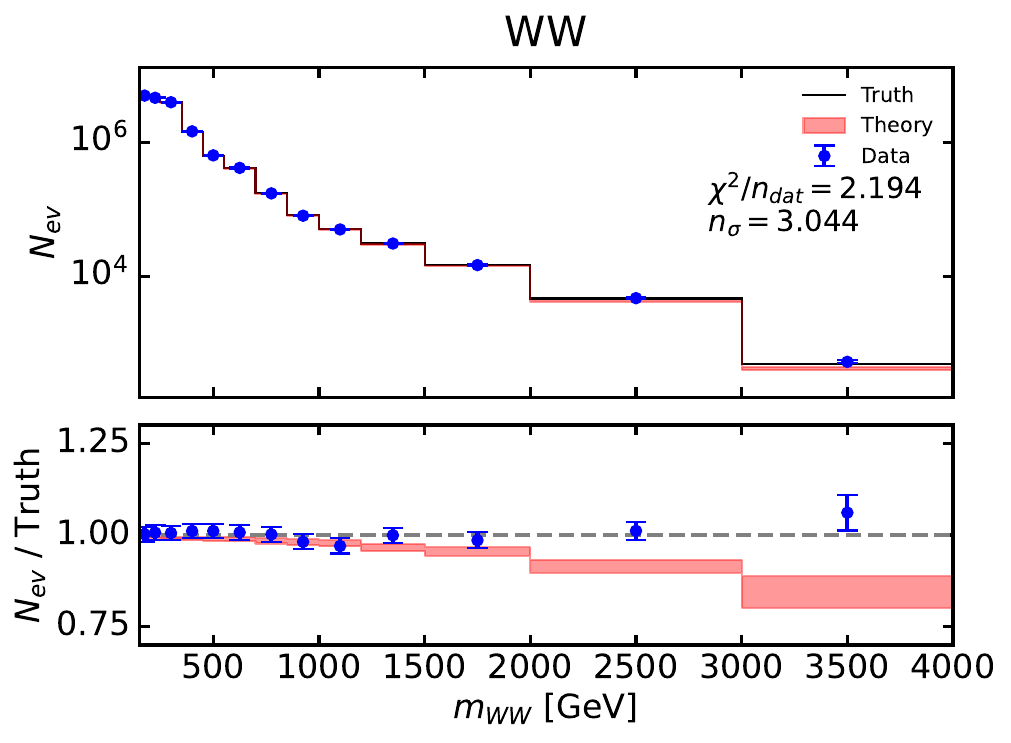}
    \caption{Predictions for $W^+ H$ (on the left) and $W^+ W^-$  (on the right) at the HL-LHC using the initial PDFs (truth) and the contaminated ones (Theory).}
    \label{fig:fake_deviations}
\end{figure}

On top of missing the new physics, the PDF contamination has another consequence. The contaminated PDFs are not compatible with the initial PDFs. Thus, they might produce non-physical discrepancies if used to compute theory predictions, even outside the DY sector. In Fig. \ref{fig:fake_deviations} the theory predictions for diboson production processes are plotted and one can observe a systematic tension between the predictions made with the initial PDFs and the contaminated ones. Theses shifts are completely fictitious and entirely caused by the fact that the PDFs have been warped by the contamination.

\section{Possible solutions to prevent contamination}

\subsection{Ratios of observables}

\begin{figure}[h!]
    \centering
    \includegraphics[width=0.49\linewidth]{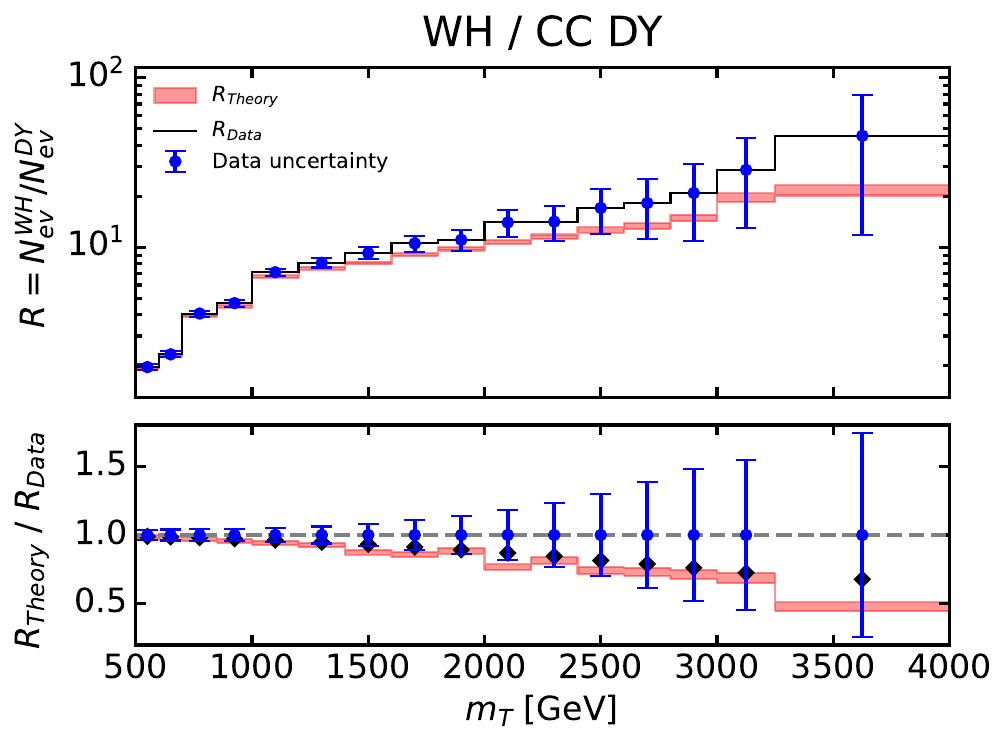}
    \includegraphics[width=0.49\linewidth]{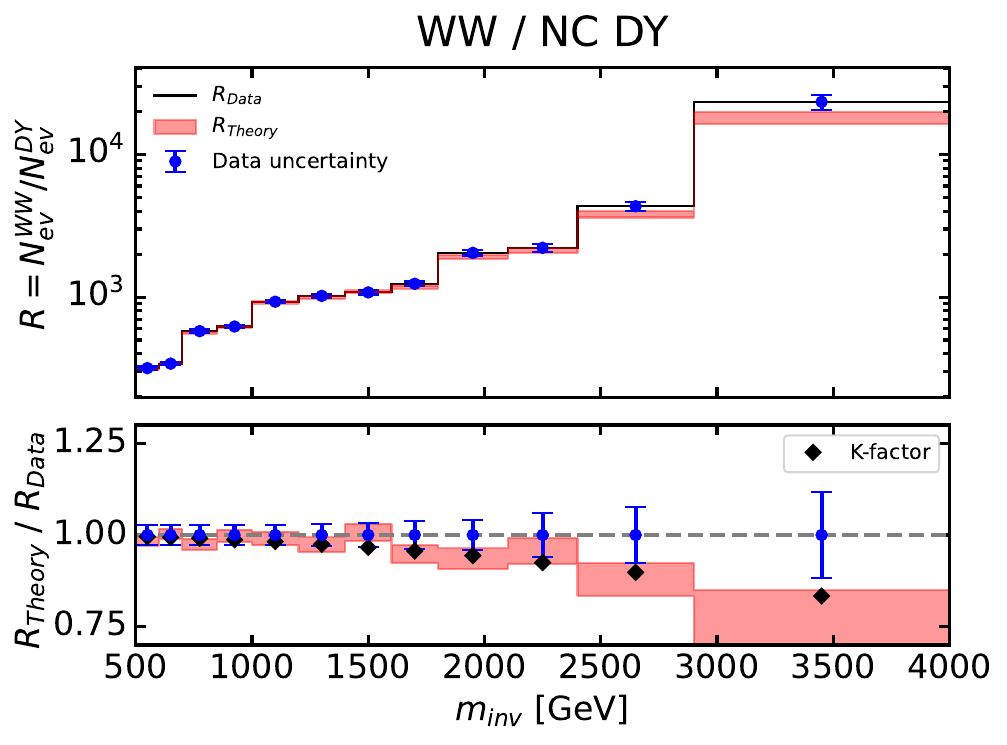}
    \caption{Top panels: Ratios of the number of events for diboson production over DY theory predictions (red) and data (black). Bottom panels: Ratio of theory predictions ratio over data ratio, compared with $K$-factors (ratio of BSM DY over SM DY cross sections).}
    \label{fig:obs_ratio}
\end{figure}

A first obvious way to discriminate possible BSM contamination in a PDF fit is to consider two different processes which share the same parton channels. By taking the ratio of those observables, the importance of the PDF is greatly diminished. Any shift from SM predictions can then be attributed to one of the partonic cross-section. In Fig. \ref{fig:obs_ratio}, one can see the ratio of the diboson over the DY cross sections.

One can observe a systematic deviation growing with the energy. This suggests the presence of new physics in one of the observables. Indeed, we know that the DY data has been generated with a theory featuring a new heavy $W'$. Without this prior knowledge, it would not be clear which of the two datasets is affected by the BSM signals. In this case only the DY pseudodata is included in the PDF fit and excluding it would effectively prevent the contamination.

\subsection{Constraints from low-energy datasets on large-x sea quarks}

Another option would be to reduce the degree of freedom available to the PDFs that allows them to adapt to the BSM signals. Practically, this corresponds to adding to the fit new datasets which would constrain the large-x sea quarks, where the PDF uncertainties are large. To be safe from heavy new physics contamination, one should consider low-energy observables. In our study we had included data from the fixed target DY experiment SeaQuest \cite{SEA}. The contamination of the PDF worsened the $\chi ^2$ function measuring the data-theory agreement for this experiment but not sufficiently to impact the fit and excluding the contaminated DY data.
The strategy would be to increase the amount of data mapping this region. The EIC programme for instance will produce important inputs in this parameter space \cite{RK} \cite{RA}.

\end{document}